\documentclass[aps,prb,reprint,a4paper,floatfix,amsmath,amssymb,amsfonts]{revtex4-1}
\usepackage{mathptmx}
\usepackage[scaled=0.9]{helvet}
\usepackage[utf8]{inputenc}
\usepackage{graphicx}
\usepackage{textcomp}
\usepackage{upgreek} 
\usepackage[dvipsnames]{xcolor}
\usepackage{soul} 
\usepackage{xspace}
\usepackage[rightcaption]{sidecap}
\usepackage{units}

\usepackage[
,textwidth=17.5cm
,textheight=23.5cm
,verbose
,pdftex
]{geometry}

\DeclareMathAlphabet{\mathcal}{OMS}{cmsy}{m}{n} 

\setcitestyle{numbers,square}

\begin{document}

\makeatletter
\renewcommand\@biblabel[1]{\mbox{$\left[ \right.$}#1\mbox{$\left. \right]$}}
\makeatother


\title{Elastic vs.\ plastic strain relaxation in coalesced GaN nanowires: an x-ray diffraction
study}

\author{Vladimir M. Kaganer}
\author{Bernd Jenichen}
\author{Oliver Brandt}

\affiliation{Paul-Drude-Institut für Festkörperelektronik, Hausvogteiplatz 5--7, 10117 Berlin, Germany}

\date{\today}
\begin{abstract}
The coalescence in dense arrays of spontaneously formed GaN nanowires proceeds by bundling: adjacent nanowires bend and merge at their top, thus reducing their surface energy at the expense of the elastic energy of bending. We give a theoretical description of the energetics of this bundling process. The bending energy is shown to be substantially reduced by the creation of dislocations at the coalescence joints. A comparison of experimental and calculated x-ray diffraction profiles from ensembles of bundled nanowires demonstrates that a large part of the bending energy is indeed relaxed by plastic deformation. The residual bending manifests itself by extended tails of the diffraction profiles.
\end{abstract}
\maketitle

\section{Introduction}

Nanowires are a natural mode of GaN growth in plasma-assisted molecular beam epitaxy, when effectively N rich conditions and sufficiently high temperatures are provided \cite{garcia98,garrido09,garrido13}. GaN nanowires then form spontaneously on various substrates without the need of any metal droplets that are required for many other semiconductor materials to induce the vapor-liquid-solid growth of one-dimensional nanostructures \cite{wagner64,wacaser09,dubrovskii15chapter}. One of the distinct advantages of the spontaneous formation and subsequent uniaxial growth of GaN nanowires is the possibility to realize abrupt axial heterojunctions between different III-N compounds by simply switching the group III supply \cite{armitage10,chang10,rigutti10,hille14,beeler15,kioseoglou15}.

Spontaneously grown GaN nanowires usually form dense arrays. The high nanowire density is advantageous for light emitting and energy harvesting devices as well as for sensing applications \cite{teubert11,li12,wallys12,howell13}. On the other hand, GaN nanowires coalesce during growth because of their proximity \cite{brandt14}. Recently, we have identified the origin of this coalescence in GaN nanowire ensembles \cite{kaganer16bundling}. Dense ensembles of GaN nanowires evolve during growth to ensembles of nanowires with smaller densities and larger diameters. However, the fraction of the surface area covered by the nanowires remains unchanged, which rules out radial growth of nanowires as the origin of coalescence. Instead, we have found that the nanowires bundle together after reaching a certain critical height to reduce the surface energy of their side facets at the expense of the elastic energy due to their bending \cite{kaganer16bundling}.

In the present work, we show that the elastic bending energy can be reduced by plastic relaxation at the coalescence joints. Dislocations at the joints \cite{consonni09,jenichen11,grossklaus13,fan14} form small angle boundaries and reduce the curvature of the nanowires. Since these dislocations run across the nanowires, the strain relaxation at the free side surface restricts the range of their elastic fields to distances comparable with the nanowire diameters. The dislocation energy is comparatively small, making the introduction of dislocations energetically favorable. The calculation of the bundling energetics shows that a major part of the bending energy is expected to be released due to formation of dislocation arrays at the joints.

We study the residual nanowire bending by a line shape analysis of x-ray diffraction profiles. The bent segments of the nanowires constitute a significantly larger volume than the coalescence joints themselves and manifest themselves by extended tails of the diffraction lines. The diffraction profiles from the ensemble of bent nanowires are calculated taking into account the distributions of lengths, diameters, and distances between bundled nanowires, as well as the mutual nanowire misorientation and the divergence of the x-ray beams. With this model, we revisit our previous x-ray diffraction measurements \cite{kaganer12NWstrain} and show that the diffraction profiles are well described by the residual bending of nanowires, and that a major part of the initial bending is released by plastic deformation.

The x-ray diffraction intensity from a bent nanowire is a result of interference of the waves scattered along its entire length. As a result, the common assumption that each small volume of the sample contributes to the diffraction intensity independently and only according to its local strain state \cite{StokesWilson44,WilliamsonHall53} is not applicable, and the dependence of the integral breadth on the reflection order does not follow a linear law.

\section{Energetics of nanowire bundling}
\label{sec:Energetics}

\subsection{Elastic relaxation}

\label{sub:ElasticRelax}

\begin{figure}
\includegraphics[width=1\columnwidth]{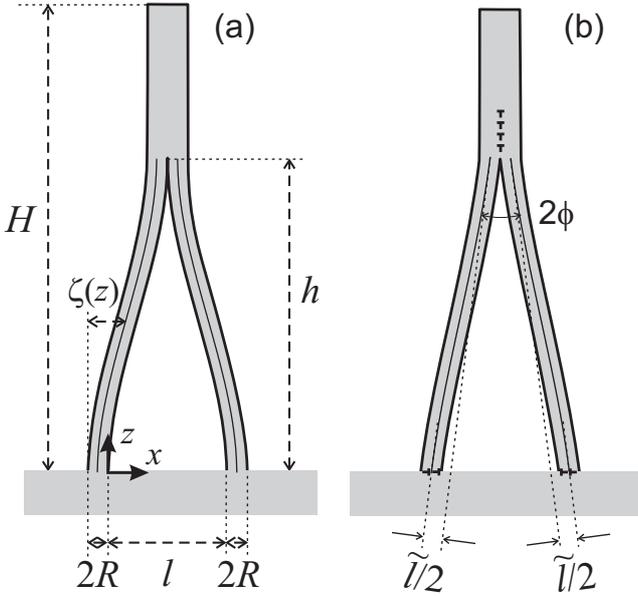} \protect\caption{Bundling of two primordial nanowires. (a) Purely elastic bending and (b) plastic relief of the bend-induced strain by creation of a dislocation array at the joint. The effective length $\tilde l$ describes the residual nanowire bending.}
\label{fig:sketch}
\end{figure}

Let us consider first the energetics of a pair of bundled nanowires that are bent to bridge their mutual separation and remain elastically strained, as shown in Fig.~\ref{fig:sketch}(a). The following calculation is based on our previous study \cite{kaganer16bundling} and serves as a reference for the analysis of plastic deformation at the joints in the next section.

We treat nanowires as thin rods experiencing small deflection (Ref.~\onlinecite{landau:elasticity}, \textsection 20). The rod is described by the deflection amplitude $\zeta(z)$, where $z$ is the coordinate along the rod. The deflection amplitude satisfies, in the absence of a shearing force, an elastic equilibrium equation $\zeta''''=0$, where the prime denotes the derivative over the coordinate $z$ along the rod. The origin is taken at the substrate surface where the bottom end of the nanowire is clamped. Hence, one has $\zeta\left|_{z=0}\right.=0$ and $\zeta'\left|_{z=0}\right.=0$.  Let the nanowires bundle at a height $h$ and touch each other for $h<z<H$. Then, the second pair of the boundary conditions at the junction is $\zeta\left|_{z=h}\right.=l/2$ and $\zeta'\left|_{z=h}\right.=0$.  Here $l$ is the distance between the nanowire surfaces at $z=0$ {[}see Fig.~\ref{fig:sketch}(a){]}. The solution of the elastic equilibrium equation with these boundary conditions is
\begin{equation} \zeta=\frac{l}{2}\left(\frac{3z^{2}}{h^{2}}-\frac{2z^{3}}{h^{3}}\right).\label{eq:1}
\end{equation}

The elastic energy of the bent nanowire is equal to (Ref.~\onlinecite{landau:elasticity}, \textsection 18)
\begin{equation} \mathcal{E}_{\mathrm{bend}}=\frac{EI}{2}\int_{0}^{h}\left(\zeta''\right)^{2}\, dz=\frac{3l^{2}EI}{2h^{3}},\label{eq:2}
\end{equation} where $E$ is the Young modulus and $I$ is the geometrical moment of inertia of the cross section of the rod. For a circular cylinder of radius $R$, the moment of inertia is equal to
\begin{equation} I=\frac{\pi}{4}R^{4}.\label{eq:3}
\end{equation} If the cross-section of the cylinder is a regular hexagon of side length $a$, the moment of inertia is
\begin{equation} I=\frac{5\sqrt{3}}{16}a^{4}.\label{eq:3a}
\end{equation}

The surface energy gained by each nanowire due to the contact of nanowires at $h<z<H$ is equal to
\begin{equation} 
\mathcal{E}_{\mathrm{surf}}(h)=-\gamma w(H-h),\label{eq:4}
\end{equation} 
where $\gamma$ is the surface energy and $w$ is the width of the contact area. For hexagonal nanowires, $w$ is equal to the side length $a$, while for a circular cylinder it can be taken equal to the radius $R$.

The nanowires bundle if the total energy $\mathcal{E}=\mathcal{E}_{\mathrm{bend}}+\mathcal{E}_{\mathrm{surf}}$ reduces. Our aim now is to find, for a given distance $l$ between nanowire sidewalls, the critical nanowire length $H_{c}$ such that the bundling is energetically favorable for $H>H_{c}$. The minimum of the total energy is given by the condition $d\mathcal{E}/dh=0$, and energy is gained by bundling if, at the minimum, $\mathcal{E}<0$.  The critical value $h_{c}$ is thus determined from the simultaneous solution of the equations $\mathcal{E}(h)=0$ and $d\mathcal{E}(h)/dh=0$.  Using expressions (\ref{eq:2}) and (\ref{eq:4}) for the energies and dividing one equation by the other, we find $h_{c}=3H_{c}/4$.  Hence, at the moment when the bundling becomes energetically favorable, one quarter of the total nanowire lengths merge together. Further straightforward calculation yields
\begin{equation} 
h_{c}=\left(\frac{9Il^{2}}{2\Lambda w}\right)^{1/4},\label{eq:5}
\end{equation} 
where $\Lambda=\gamma/E$. 
With the Young modulus of GaN $E=3.55\times10^{11}$~J/m$^{3}$ 
and a surface energy of $\gamma=118$~meV/$\mathrm{\AA}^{2}=1.9$~J/m$^{2}$ 
for the \emph{M}-plane facets that constitute the GaN nanowire sidewalls \cite{northrup96}, we get
$\Lambda=0.53\times10^{-2}$~nm.

For an estimate, consider two hexagonal nanowires with side facet width $a=10$~nm at a distance $l=30$~nm between their side facets.  The distance between the nanowire centers is then $l+2a=50$~nm, which corresponds to the mean distance between nanowires for a density of 4$\times10^{10}$~cm$^{-2}$, the typical density of nucleated nanowires before the bundling starts \cite{kaganer16bundling}. The critical height given by Eq.~(\ref{eq:5}) is $h_{c}=143$~nm. The total length of the nanowires is $H_{c}=4h_{c}/3=190$~nm. The elastic energy (\ref{eq:2}) stored in each of the bent pillars of the critical height ($h=h_{c})$ is equal to
\begin{equation} \mathcal{E}_{\mathrm{bend}}=E\left(\frac{\Lambda^{3}w^{3}Il^{2}}{18}\right)^{1/4}.\label{eq:6}
\end{equation} For the example above, this energy is $\mathcal{E}_{\mathrm{bend}}=8.9\times10^{-16}$~J.

Bundeled nanowires may experience subsequent bundling with other nanowires in their vicinity.  The energetics of this process can, in principle, be described in the same way as above. However, the bending energy of nanowire bundles staying on several pillars can only be evaluated numerically. In our previous work \cite{kaganer16bundling}, we have approximated this situation by replacing the bundled pair by a single cylindrical nanowire with a volume and cross-sectional area equal to the respective sums of the constituents.

\subsection{Plastic relaxation via dislocation network}

\subsubsection{Energy of the dislocation network}

\label{sub:DislEnergy}

The energy of bending can be partially or completely released if dislocations are present at the joint and form a small angle boundary, as shown in Fig.~\ref{fig:sketch}(b). The release of the elastic energy of bending costs an extra energy of dislocations. Let us estimate the dislocation energy first.

The energy of a straight dislocation is equal to \cite{hirthlothe82}
\begin{equation} E_{d}=\frac{\mu b^{2}w}{4\pi}\ln\frac{D}{b}.\label{eq:7}
\end{equation} Here $\mu$ is the shear modulus, $b$ is the Burgers vector, the length of the dislocation is taken equal to the width $w$ of the contact area introduced above, and $D$ is a cutoff distance. For a dislocation running across the nanowire, as shown in Fig.~\ref{fig:sketch}(b), the dislocation strain field is relaxed at distances exceeding the nanowire diameter, and for this reason the cutoff distance in Eq.~(\ref{eq:7}) is taken to be equal to the nanowire diameter $D$. Since the dislocation energy depends on the cutoff distance only logarithmically, this choice is not critical.

A row of dislocations with a separation $p$ provides a relative rotation of the two nanowires by the angle $2\phi=b/p$. Since the dislocation strain is relaxed on distances exceeding the nanowire diameter, dislocations located at distances from the joint exceeding the diameter would have only little effect on the rotation angle $\phi$. Hence, the total length of the row can be taken equal to $D$. This situation contrasts to the common case of relative rotation of two bulk crystallites where the dislocation row runs along the whole boundary between crystallites.  Denoting the number of the dislocations in the row by $N$, we have $p=D/N$ and hence
\begin{equation} \phi=bN/2D.\label{eq:8}
\end{equation}

If the dislocations in the row are equidistant, the dislocation strain is relaxed on the distance $p$ and the logarithmic term in Eq.~(\ref{eq:7}) is equal to $\ln(p/b)$. However, we keep the logarithmic term as it is written in Eq.~(\ref{eq:7}), thus accounting for a disorder in the dislocation positions. The difference between these terms is in any case minor. We express the shear modulus via the Young modulus, $\mu=E/2(1+\nu)$, where $\nu$ is the Poisson ratio, and write the energy of the dislocation row at the joint in Fig.~\ref{fig:sketch}(b) as
\begin{equation} \mathcal{E}_{\mathrm{disl}}=N\frac{Eb^{2}w}{8\pi}\mathcal{L},\label{eq:9}
\end{equation} where $\mathcal{L}=(1+\nu)^{-1}\ln(D/b)$. Taking the Burgers vector equal to the in-plane lattice parameter of GaN, $b=0.32$~nm, the nanowire diameter $D=20$~nm, and the Poisson ratio $\nu=0.2$, we get $\mathcal{L}\approx3.4$. We use this value in all calculations below, taking into account that a variation of the nanowire diameter changes this factor only very little because of the logarithmic dependence.

The atomic arrangement at the boundary between the nanowire and the substrate is unknown. Prior to nanowire nucleation, the Si substrate is nitridated by the N plasma inducing the formation of an amorphous SiN$_{x}$ interlayer. Nevertheless, the in-plane orientation of the nanowires clearly follows that of the Si(111) substrate with a deviation (twist) of about 3$^{\circ}$ \cite{jenichen11,wierzbicka13}. To simplify the calculations below and to preserve a symmetric shape of the nanowires, we assume that relaxation at the interface proceeds in the same way as at the joints. Hence, we assume that the interface between each nanowire in Fig.~\ref{fig:sketch}(b) and the substrate contains $N/2$ dislocations with the Burgers vectors of the same length $b$ (Burgers vectors are directed vertically in this case), so that each nanowire is inclined by the same angle $\phi$ given by Eq.~(\ref{eq:8}). The total energy of all dislocation arrays in Fig.~\ref{fig:sketch}(b) is thus twice the energy given by Eq.~(\ref{eq:9}).  Finally, the total dislocation energy per nanowire is described by Eq.~(\ref{eq:9}).

Let us calculate the energy of the dislocations required to totally relax the elastic bending in the same example as considered at the end of the previous section, $a=10$~nm and $l=30$~nm. With the length $h=143$~nm obtained there, a rotation of the pillars by the angle $\phi=\arctan(l/2h)\approx6{}^{\circ}$ requires, according to Eq.~(\ref{eq:8}), $N=13$ dislocations, whose total energy (\ref{eq:9}) is equal to $\mathcal{E}_{\mathrm{disl}}=6.6\times10^{-16}$~J, smaller than the elastic energy of bending (\ref{eq:6}) calculated above for the same parameters. This example shows that the bending energy (\ref{eq:6}) and the dislocation energy (\ref{eq:9}) are comparable, and the relaxation by creation of dislocations at the nanowire joints may be energetically favorable. In the next section, we generalize our analysis of the energy balance by considering a partial relaxation by dislocations.

Let us now estimate the total density of dislocations introduced into the nanowire ensemble. For an array of parallel dislocations in a bulk crystal, the dislocation density is defined as the number of dislocations per unit area oriented normally to the dislocation lines. A more general definition of the dislocation density is the total length of the dislocation lines per unit volume. We calculate the dislocation density using this latter definition and ascribing $N=13$ dislocations of length $w=10$~nm to the volume of two nanowires of hexagonal shape with side lengths $a=10$~nm and heights $2h=286$~nm. The dislocation density thus calculated is $9\times10^{9}$~cm$^{-2}$, comparable to the density of threading dislocations in heteroepitaxial GaN layers grown by molecular beam epitaxy. However, one has to keep in mind that these dislocations do not propagate along the nanowire axis, but are localized at the joints. In addition, their strain fields are relaxed at distances from the joints exceeding the nanowire diameter. As a result, the impact of these dislocations on, particularly, the luminous efficiency of nanowires is certainly much less pronounced than the effect of the same density of dislocations intersecting a heteroepitaxial layer.

\subsubsection{Energetics of plastic relaxation\label{sub:PlasticRelax}}

Let us assume that $N$ dislocations are created at the nanowire joint in Fig.~\ref{fig:sketch}(b). Dislocations are distributed over an interval $D$ and provide an angle between the two pillars at the joint $2\phi=Nb/D$. We assume, as discussed above , that at the interface between each nanowire and the substrate, $N/2$ dislocations are created and induce the inclination of the nanowires by an angle $\phi$ in direction to each other. 

The same elastic equilibrium equation $\zeta''''=0$ as in Sec.~\ref{sub:ElasticRelax} is to be solved now with another set of boundary conditions. We have $\zeta\left|_{z=0}\right.=0$ and $\zeta'\left|_{z=0}\right.=\phi$ at the substrate and $\zeta\left|_{z=h}\right.=l/2$, $\zeta'\left|_{z=h}\right.=\phi$ at the joint. The solution is
\begin{equation} 
\zeta=\frac{\tilde{l}}{2}\left(\frac{3z^{2}}{h^{2}}-\frac{2z^{3}}{h^{3}}\right)+\phi z,\label{eq:10}
\end{equation} 
generalizing Eq.~(\ref{eq:1}) by replacing $l$ with the residual quantity
\begin{equation} 
\tilde{l}=l-2\phi h.
\label{eq:11}
\end{equation}
The effective distance $\tilde l$ describes the residual bending of nanowire segments.
It is always smaller than the actual distance $l$ between the nanowires and approaches
zero for complete relaxation. 
For the elastic energy of the residual nanowire bending we have, instead of Eq.~(\ref{eq:2}),
\begin{equation} 
\mathcal{E}_{\mathrm{bend}}=\frac{3\tilde{l}{}^{2}EI}{2h^{3}}.
\label{eq:12}
\end{equation}

Our aim now is to find the minimum of the total energy $\mathcal{E}=\mathcal{E}_{\mathrm{bend}}+\mathcal{E}_{\mathrm{surf}}+\mathcal{E}_{\mathrm{disl}}$ with respect to the number of dislocations $N$ and the nanowire length $h$. Since $\mathcal{E}_{\mathrm{surf}}$ does not depend on $N$, we find first the minimum of the sum of $\mathcal{E}_{\mathrm{bend}}$, described by Eq.~(\ref{eq:12}), and $\mathcal{E}_{\mathrm{disl}}$, given by Eq.~(\ref{eq:9}), over $N$. The number of dislocations is considered as a continuous variable. The minimum over $N$ is reached at
\begin{equation} N=\frac{lD}{bh}-\frac{\mathcal{L}}{24\pi}\frac{D^{2}hw}{I}\label{eq:13}
\end{equation} and is equal to
\begin{equation} \mathcal{E}_{\mathrm{bend}}+\mathcal{E}_{\mathrm{disl}}=E\left(\frac{lv}{h}-\frac{hv^{2}}{6I}\right),\label{eq:14}
\end{equation} where
\begin{equation} v=\frac{\mathcal{L}}{8\pi}bwD.\label{eq:15a}
\end{equation}

Using Eqs.~(\ref{eq:8}), (\ref{eq:11}), and (\ref{eq:13}), we can represent the effective length describing the remaining bending of nanowires as
\begin{equation} \tilde{l}=\frac{h^{2}v}{3I},\label{eq:19}
\end{equation} while the number of dislocations (\ref{eq:13}) providing the minimum of energy can be written as
\begin{equation} 
N=\frac{(l-\tilde{l})D}{bh}.
\label{eq:19a}
\end{equation} 
The condition of the energy gain by introduction of dislocations $N>0$ can be written now as $\tilde{l}<l$.

The minimum of the sum of Eq.~(\ref{eq:14}) and the surface energy (\ref{eq:4}) over $h$ is
\begin{equation} \mathcal{E}_{\mathrm{bend}}+\mathcal{E}_{\mathrm{disl}}+\mathcal{E}_{\mathrm{surf}}=\gamma w(H'_{c}-H),\label{eq:15}
\end{equation} where it is denoted
\begin{equation} 
H'_{c}=2\left(\frac{lv\xi}{\Lambda w}\right)^{1/2}\label{eq:16}
\end{equation} 
and
\begin{equation} 
\xi=1-\frac{v^{2}}{6\Lambda wI}.\label{eq:17}
\end{equation} With the values of the parameters used in the example above, we find $1-\xi=0.044$. Hence, $\xi\approx1$ in all practical cases. The minimum of the total energy (\ref{eq:15}) is reached at
\begin{equation} 
h=\frac{H'_{c}}{2\xi}.\label{eq:18}
\end{equation} 
Thus, the gain in surface energy due to bundling exceeds the sum of the bending energy and the dislocation energy for $H>H'_{c}$, and the length $h$ of the two separate nanowires below the joint is approximately $H'_{c}/2$.

Let us make an estimate for the same values of parameters as in Sec.~\ref{sub:ElasticRelax}, namely, the side facet width $a=10$~nm and the distance $l=30$~nm.  The critical length calculated by Eq.~(\ref{eq:16}) is $H'_{c}=138$~nm, smaller than the the critical length for purely elastic relaxation in the absence of dislocations. The joint is at the height $h=72$~nm.  For the effective distance $\tilde{l}$ describing the residual curvature of the nanowires, Eq.~(\ref{eq:19}) gives $\tilde{l}\approx2.6$~nm, an order of magnitude smaller than the real distance $l$ between nanowires. The number of dislocations (\ref{eq:19a}) providing this minimum is $N=24$. The number of dislocations obtained in this estimate is almost by a factor of two larger than the number of dislocations calculated in the example of a complete plastic relaxation in the previous section, since the length of the bundling pillars is two times smaller now, and the angle $2\phi$ is correspondingly larger. Dislocations release most of the nanowire bending, leaving the effective distance $\tilde l$, which describes the residual bending, much smaller than the distance $l$ between nanowires. Thus, most of the elastic energy of bending can be released by creation of dislocations at the joint.

The energy minimization gives a very dense dislocation array, with the distance between dislocations of about 1~nm. However, plastic relaxation at a joint to the energy minimum may be restricted. At the beginning of a bundling event, two nanowires first touch each other by atomically flat side facets. An introduction of dislocations at a border between two perfect crystallites may be kinetically suppressed, despite being energetically favorable. If, however, a sufficient number of dislocations is introduced to minimize the energy, the distances between dislocations are fairly small, since all dislocations that relax the relative misorientation of nanowires need to be located at distances from the joint not exceeding the nanowire diameter. They have to relax rather large relative misorientations of nanowires. The elastic energy of such a dense dislocation array may be reduced further by building a large-angle boundary with a coincidence site lattice at the interface. Transmission electron microscopy studies of GaN nanowires reveal dislocations at the coalescence joints \cite{consonni09,jenichen11,fan14} and also ``zipper'' defects \cite{grossklaus13} whose atomic structure remains obscure. Hence, further microscopic studies are needed to understand the relaxation at the nanowire joints in more detail.

\subsubsection{Relaxation of the nanowire twist}

\begin{figure}
\includegraphics[width=1\columnwidth]{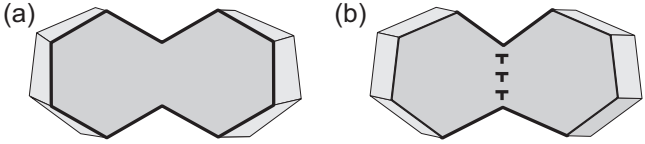}

\protect\caption{Relaxation of the in-plane nanowire misorientation by (a) elastic twist deformation and (b) a dislocation array.}
\label{fig:twist}
\end{figure}

The in-plane relative misorientation of the nanowires (twist) can be released upon bundling either elastically by a torsion deformation of the nanowires, as shown in Fig.~\ref{fig:twist}(a), or plastically by creation of an array of dislocations, as depicted in Fig.~\ref{fig:twist}(b).  Dislocations created in this case are threading dislocations running along the nanowires. During further growth, these threading dislocations are likely to emerge from the nanowire at its sidewall. In this section, we estimate the energy of the torsion deformation.

The torsion energy of a thin rod is (Ref.~\onlinecite{landau:elasticity}, \textsection 16)
\begin{equation} \mathcal{E}_{\mathrm{tors}}=Ch\tau^{2}/2,\label{eq:26}
\end{equation} where $h$ is, as above, the length of the nanowire up to the joint, $C$ is the torsional rigidity, and $\tau$ is the torsion angle defined as the angle of rotation per unit length. The torsional rigidity for a rod with the circular cross section is (Ref.~\onlinecite{landau:elasticity}, \textsection 16, problem 1) $C=\mu\pi R^{4}/2$. The torsion angle is $\tau=\phi_{\mathrm{twist}}/h$, where $2\phi_{\mathrm{twist}}$ is the mutual in-plane misorientation of the two nanowires. Expressing the shear modulus $\mu$ via the Young modulus, $\mu=E/2(1+\nu)$, we rewrite Eq.~(\ref{eq:26}) as
\begin{equation} 
\mathcal{E}_{\mathrm{tors}}=\frac{EI}{2(1+\nu)}\frac{\phi_{\mathrm{twist}}^{2}}{h},
\label{eq:27}
\end{equation} where $I$ is the moment of inertia of the circle of radius $R$ (\ref{eq:3}).

The elastic energy of bending (\ref{eq:2}) can be rewritten as
\begin{equation} \mathcal{E}_{\mathrm{bend}}=6EI\frac{\phi_{\mathrm{tilt}}^{2}}{h},\label{eq:28}
\end{equation} where $\phi_{\mathrm{tilt}}=l/2h$ is the inclination angle of the nanowire due to bundling, as shown in Fig.~\ref{fig:sketch}(b).  This angle is denoted simply by $\phi$ in the rest of the paper.  In this section, it is referred to as $\phi_{\mathrm{tilt}}$, to clearly distinguish it from the twist angle. The ratio of the torsion to the bending energy is
\begin{equation} \frac{\mathcal{\mathcal{E}_{\mathrm{tors}}}}{\mathcal{E}_{\mathrm{bend}}}=\frac{1}{12(1+\nu)}\left(\frac{\phi_{\mathrm{twist}}}{\phi_{\mathrm{tilt}}}\right)^{2}.\label{eq:29}
\end{equation}
Taking the tilt and the twist angles to be equal, one can see that Eq.~(\ref{eq:29}) contains a small numerical factor $[12(1+\nu)]^{-1}\approx0.07$ which makes creation of threading dislocations to relax torsion energy less favorable compared with creation of dislocations running across the nanowires that relax the nanowire bending energy. We do not have experimental evidences of the torsion relaxation that would stimulate further analysis.  The x-ray diffraction study presented below is not sensitive to torsion, since only symmetric Bragg reflections are analyzed. Therefore, we restrict ourselves to the consequences of nanowire bending as analyzed in Sec.\ \ref{sub:PlasticRelax}.

\section{X-ray diffraction from bundled nanowires}
\label{sec:diffraction}

\subsection{Diffraction from a single nanowire}

Let us calculate the x-ray diffraction intensity from a single distorted nanowire whose shape is described by Eq.~(\ref{eq:10}). We restrict our study to symmetric Bragg reflections and consider the displacement of atomic planes in $z$ direction,
\begin{equation} u_{z}=x\frac{\partial\zeta}{\partial z}.\label{eq:20}
\end{equation} The scattering vector can be written as $\mathbf{Q}_{0}+\mathbf{q}$, where $\mathbf{Q}_{0}$ is the reciprocal lattice vector (directed along the $z$ axis) and $\mathbf{q}$ is a small deviation from it. We calculate here the scattering intensity $I(\mathbf{q})$ for an arbitrary direction of $\mathbf{q}$, with the aim to perform an integration over orientations of nanowires and the x-ray beam divergence in the next section. The scattering amplitude is an integral over the volume of the nanowire,
\begin{eqnarray} A(\mathbf{q}) & = & \int\exp(i\mathbf{q}\cdot\mathbf{r}+i\mathbf{Q}_{0}\cdot\mathbf{u})d\mathbf{r}\label{eq:21}\\ & = & \intop_{0}^{h}\negthickspace dz\intop_{-R}^{R}\negthickspace dx\intop_{-R}^{R}\negthickspace dy\,\exp(iq_{x}x+iq_{y}y+iq_{z}z+iQ_{0}u_{z}).\nonumber
\end{eqnarray}

The nanowire cross section is considered here, for the sake of simplicity, to be a square with sides $2R$. A straightforward integration with the displacement field described by Eqs.~(\ref{eq:10}) and (\ref{eq:20}) gives
\begin{eqnarray} A(\mathbf{q}) & =hR & \frac{\sin q_{y}R}{q_{y}}\intop_{0}^{1}d\xi\:\cos\left(q_{z}h\,\xi/2\right)\nonumber \\ & \times & \frac{\sin\left[\kappa\left(1-\xi^{2}\right)+q_{x}R\right]}{\kappa\left(1-\xi^{2}\right)+q_{x}R},\label{eq:22}
\end{eqnarray} where
\begin{equation} \kappa=3Q_{0}\tilde{l}R/4h.\label{eq:23}
\end{equation} Particularly, if the nanowire is not bent ($\tilde{l}=0$), Eq.~(\ref{eq:22}) reduces to the common expression for the x-ray scattering amplitude from a perfect crystal of a parallelepiped of sizes $2R\times2R\times h$,
\begin{equation} A(\mathbf{q})=\frac{\sin(q_{x}R)}{q_{x}}\frac{\sin(q_{y}R)}{q_{y}}\frac{\sin(q_{z}h/2)}{q_{z}}.\label{eq:24}
\end{equation} The x-ray scattering intensity is
\begin{equation} I(\mathbf{q})=\left|A(\mathbf{q})\right|^{2}.\label{eq:25}
\end{equation}

A direct calculation of the scattering intensity by Eqs.~(\ref{eq:22}) or (\ref{eq:24}) gives rise to oscillations which are not observed in the experiment because of statistical variations of nanowire lengths, diameters, distances between bundled nanowires, as well as a distribution in the nanowire orientations and the x-ray beam divergence. In the next section, we average the intensity (\ref{eq:25}) over all these distributions.

\subsection{Diffraction from an ensemble of nanowires}

The x-ray scattering intensity $I(\mathbf{q})$ is calculated in the previous section for a single nanowire in the coordinate system that refers to its orientation. The x-ray intensity scattered from an ensemble of nanowires involves two different kinds of integration. First, the nanowires possess a range of orientations, with a typical width of few degrees. Second, the x-ray beam illuminating the sample in a laboratory experiment is well collimated in the scattering plane but has a divergence, also a few degrees, normal to this plane (the vertical divergence). As a result, a nanowire with an orientation different from the reference orientation may provide a strong scattering due to an appropriate component of the incident x-ray beam. This effect manifests itself in the asymmetric x-ray diffraction profiles from nanowire ensembles \cite{kaganer12NWstrain} and needs to be properly taken into account in the present analysis.

The scattering amplitude (\ref{eq:22}) is written in the frame defined by the nanowire orientation: the $z$ axis is along the nanowire and the $x$ axis is along the line connecting two bundled nanowires, as shown in Fig.~\ref{fig:sketch}. Our aim now is to find the components of the vector $\mathbf{q}$ in this frame. Let us consider first a distribution in orientations of the unit vectors $\mathbf{n}$ along the nanowires for a given scattering vector $\mathbf{K}$. We write identically
\begin{equation} \mathbf{Q}=(\mathbf{K}\cdot\mathbf{n})\mathbf{n}+[\mathbf{K}-(\mathbf{K}\cdot\mathbf{n})\mathbf{n}].\label{eq:30}
\end{equation} The first term is a vector along the nanowire, so that $q_{z}=\mathbf{K}\cdot\mathbf{n}-Q_{0}$.  The second term is a vector in the ($x,y$) plane perpendicular to the nanowire, so that $(q_{x},q_{y})=\mathbf{K}-(\mathbf{K}\cdot\mathbf{n})\mathbf{n}$.  The average of the intensity (\ref{eq:25}) over all nanowire orientations is the average over the distribution of orientations $\mathbf{n}$.  Since the direction of the $x$ axis has been chosen above as the direction from one nanowire to the other in the pair of bundled nanowires, as it is shown in Fig.~\ref{fig:sketch}, the average of the x-ray intensity also implies an average over the choice of the direction of $q_{x}$ axis in the ($q_{x},q_{y}$) plane.

Let us now describe the set of possible scattering vectors $\mathbf{K}$. The laboratory x-ray diffraction setup is designed to provide high resolution in the scattering plane but does not inhibit a large divergence in the direction normal to that plane. This ``vertical divergence'' gives rise to a fan of the incident beams making the same angle $\theta$ to the sample surface in projection to the reference scattering plane $(x,z)$ plane but different angles $\psi$ to that plane. The wave vectors of the incident and the scattered beams in this fan are
\begin{equation} \mathbf{K}^{\mathrm{in,out}}=\frac{2\pi}{\lambda}(\cos\theta\cos\psi,\mp\sin\psi,\mp\sin\theta\cos\psi),\label{eq:A2}
\end{equation} and the scattering vectors $\mathbf{K}=\mathbf{K}^{\mathrm{out}}-\mathbf{K}^{\mathrm{in}}$ are
\begin{equation} \mathbf{K}=\frac{4\pi}{\lambda}(0,\sin\psi,\sin\theta\cos\psi).\label{eq:A3}
\end{equation} 
Here $\lambda$ is the x-ray wavelength. Variation of the angle $\psi$ in a range corresponding to the vertical divergence of the incident beam provides a set of scattering vectors for a given scattering angle $2\theta$.

The set of waves present in the incident beam can be revealed, and the vertical divergence thus measured, by taking a perfect crystal as a sample and inclining it by an angle $\chi$ with respect to the vertical plane \cite{kaganer12NWstrain}. It is instructive to relate the inclination angle $\chi$ and the scattering angle $2\theta$ using the formalism above. When a bulk single crystal, rather than a nanowire ensemble, is used as a sample, the first two terms in Eq.~(\ref{eq:24}) reduce to $\delta(q_{x})\delta(q_{y})$, i.e., for a given inclination angle $\chi$, the scattering vector $\mathbf{K}$ is parallel to the surface normal of the sample
\begin{equation}
\mathbf{n}=(0,\sin\chi,\cos\chi). \label{eq:A31}
\end{equation}
The requirement of the vectors (\ref{eq:A3}) and (\ref{eq:A31}) to be collinear gives rise to the relation $\tan\psi=\sin\theta\tan\chi$.

The Bragg condition reads $\mathbf{K\cdot}\mathbf{n}=2\pi/d$, where $d$ is the lattice spacing for the respective reflection. Using the relation between the angles $\chi$ and $\psi$, the Bragg condition can be written, for $\chi\ll1$, as
\begin{equation} 
\sin\theta=\sin\theta_{B} \left( 1-\frac{1}{2} \chi^{2}\cos^{2}\theta_{B} \right),
\label{eq:A4}
\end{equation} where $\theta_{B}$ defined by $\sin\theta_{B}=\lambda/2d$ is the diffraction peak position at $\chi=0$. Equation (\ref{eq:A4}) describes the shift of the diffraction peak when a perfect crystal sample is inclined by an angle $\chi$. We have checked the positions of the diffraction peaks in Fig.~2 of Ref.~\onlinecite{kaganer12NWstrain} and found that they follow Eq.~(\ref{eq:A4}).

Equations (\ref{eq:30}) and (\ref{eq:A3}) are used below to calculate the x-ray diffraction intensity from an ensemble of nanowires.

\begin{figure*}
\includegraphics[width=1\textwidth]{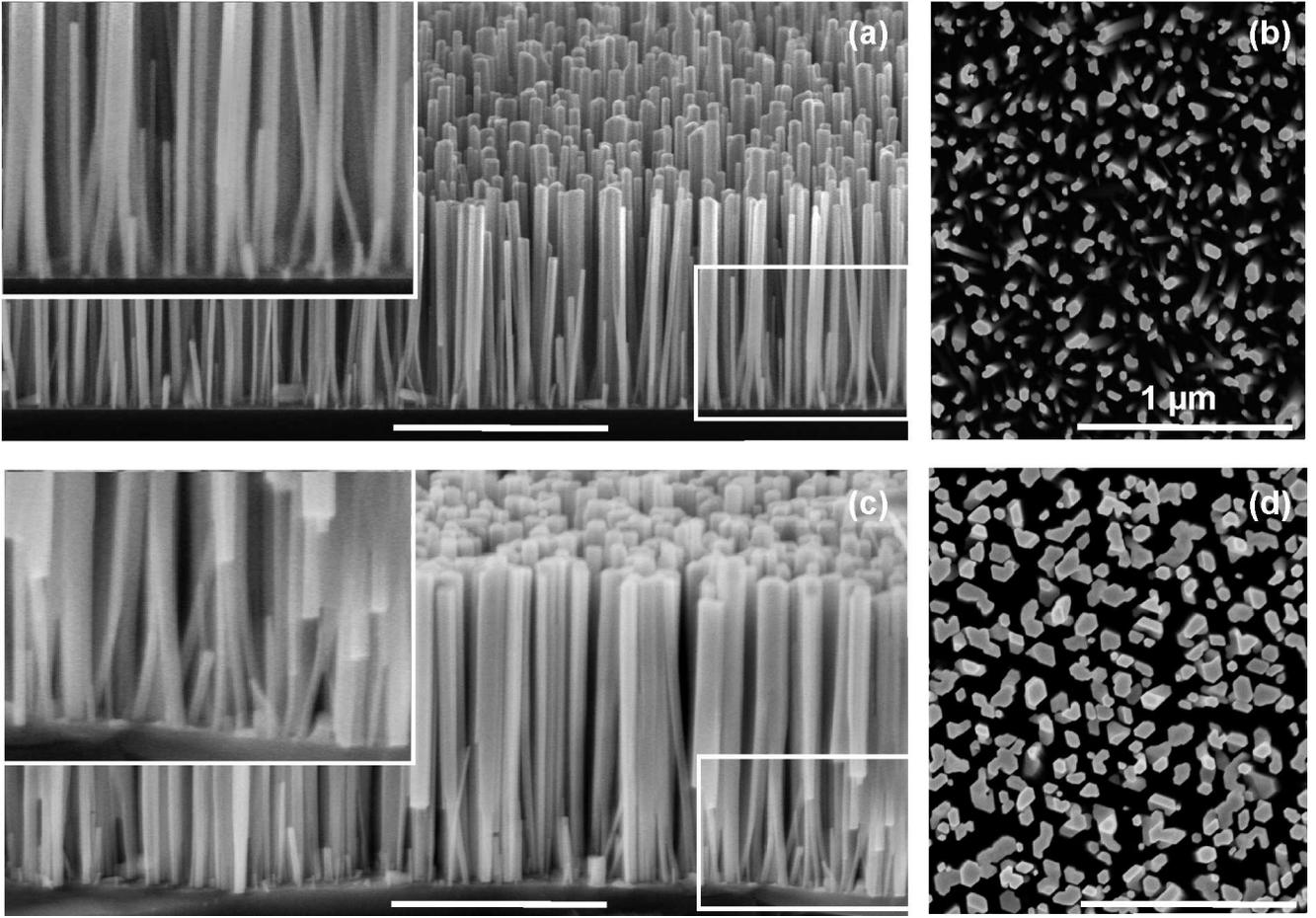}
\protect\caption{Bird's eye-view (left) and top-view (right) scanning electron micrographs of [(a), (b)] sample 1 and [(c), (d)] sample 2. Magnified parts of the bird's eye-view micrographs are shown in the top left corners of the respective images. The scale bars in all micrographs have the same length of 1 \textmu m.}
\label{fig:SEMs}
\end{figure*}

\section{X-ray diffraction line profiles}

\subsection{Experiment}

\begin{figure}
\includegraphics[width=1\columnwidth]{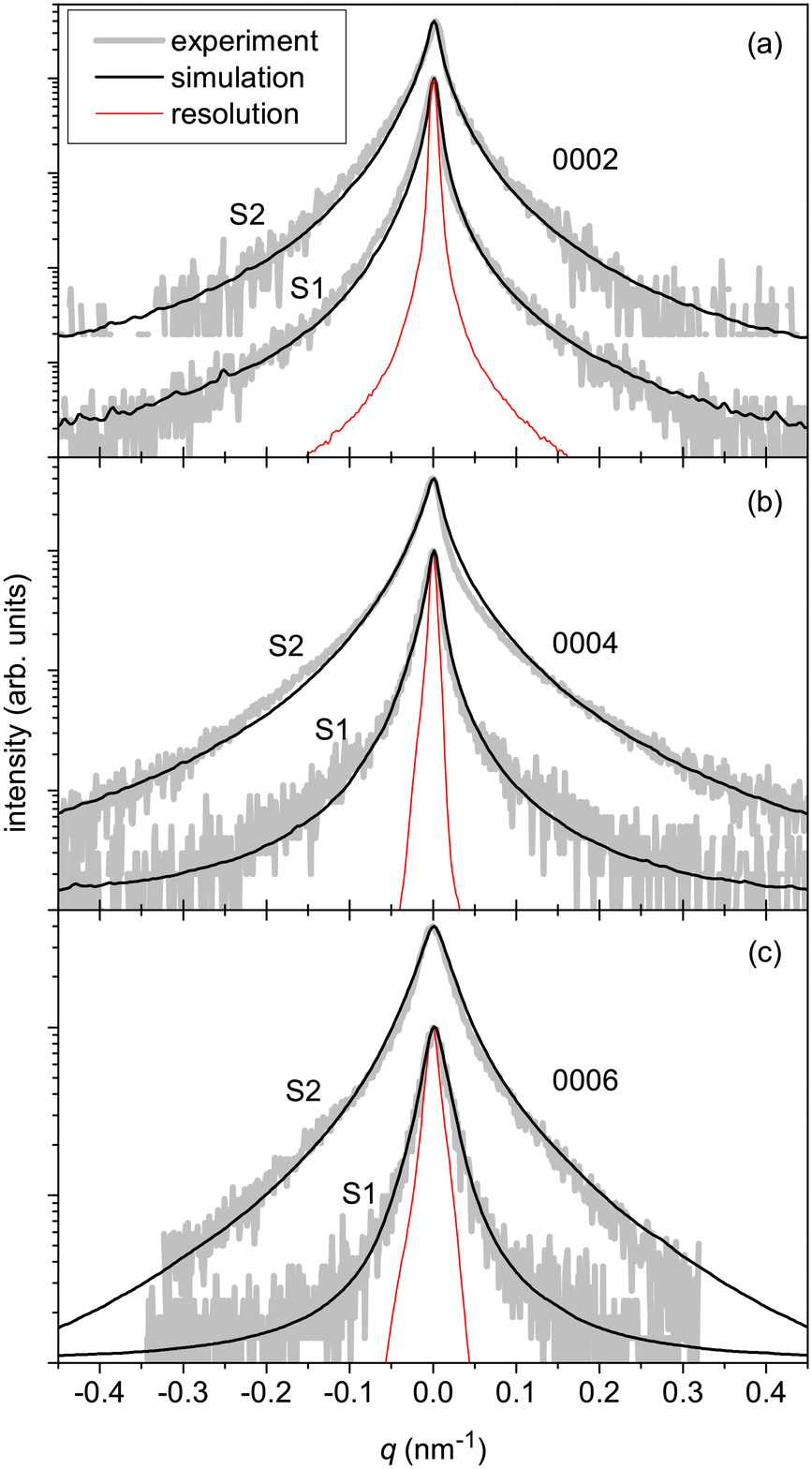}
\protect\caption{Experimental (thick gray lines) and theoretical (thin black lines) triple-crystal x-ray diffraction profiles across the (a) 0002, (b) 0004 , and (c) 0006 reflections from samples 1 and 2. The resolution functions measured for the same reflections from a bulk GaN crystal are depicted as thin red lines.}
\label{fig:XrayProfiles}
\end{figure} In the present work, we analyze anew the measurements performed in our former study \cite{kaganer12NWstrain}. The GaN nanowires were grown on Si(111) substrates by plasma-assisted molecular beam epitaxy.  Growth conditions were chosen for the synthesis of two clearly distinct samples in terms of nanowire density, average diameter, and degree of coalescence. Sample 1 (S1)  was obtained at a temperature of 810$^{\circ}$C and with a N/Ga flux ratio of 2.5, whereas for sample 2 (S2) both a lower temperature of 780$^{\circ}$C and a lower N/Ga flux ratio of 1.2 were employed. Consequently, sample 1 is characterized by a lower degree of coalescence compared to sample 2.

Representative scanning electron micrographs of both samples are shown in Fig.~\ref{fig:SEMs}. The key to the understanding of the coalescence process is gained by a close inspection of the bottom parts of the ensembles in side view \cite{kaganer16bundling} as depicted in the insets of Figs.~\ref{fig:SEMs}(a) and \ref{fig:SEMs}(c). Close to the substrate, the nanowires have diameters of 20--30~nm that they attain shortly after nucleation \cite{consonni11}. The nanowires do not grow radially in their bottom parts in the process of axial growth. Rather, they start to bundle at a height of 100--200~nm, as clearly seen in Figs.~\ref{fig:SEMs}(a) and \ref{fig:SEMs}(c) and the magnified insets. These values are in a good agreement with the numerical estimates in Sec.~\ref{sec:Energetics}.

During axial growth, the density of the nanowires continuously decreases and their diameters continuously increase, while the fraction of the area covered by nanowires remains unchanged \cite{kaganer16bundling}. This behavior is a result of a continuous bundling of nanowires into larger aggregates, a process which is visible in the top-view scanning electron micrographs in Figs.~\ref{fig:SEMs}(b) and \ref{fig:SEMs}(d) by the irregular shape of the coalesced aggregates \cite{brandt14}.

While the presence of bent nanowire segments is evident from the side-view micrographs shown in Figs.~\ref{fig:SEMs}(a) and \ref{fig:SEMs}(c), and partly also visible in the top view depicted in Figs.~\ref{fig:SEMs}(b), we cannot determine from these micrographs whether the bending is purely elastic or partly accommodated by plastic relaxation. Since x-ray diffraction is sensitive to the residual curvature of nanowires, we analyze x-ray diffraction profiles from both samples to obtain quantitative information on the elastic strain in bundled nanowires.

X-ray measurements were carried out with CuK$\alpha_{1}$ radiation using a Panalytical X'Pert diffractometer with two bounce Ge(220) hybrid monochromator and three bounce Ge(220) analyzer crystal. The symmetric 0002, 0004, and 0006 reflections were measured by $\upomega$-$2\uptheta$ Bragg scans. To obtain the resolution functions of the experimental setup at the respective scattering angles, the same reflections were also measured on a free-standing GaN layer with a dislocation density as low as 6$\times$10$^{5}$~cm$^{-2}$.

\subsection{Line profile analysis}

Figure \ref{fig:XrayProfiles} shows the measured triple-crystal x-ray diffraction profiles of samples 1 and 2 in successive orders of the symmetric Bragg reflections. The profiles measured on the free-standing GaN layer shown by thin red lines are used as resolution functions for the respective reflections. These profiles are measured at the same Bragg angles as used for the nanowire samples and thus include the chromatic aberration effects for reflections at different diffraction angles. In the present analysis, the vertical divergence of the x-ray beam and the resulting asymmetry of the diffraction profiles are taken into account in the formalism developed in Sec.\ \ref{sec:diffraction}, as described below.  Therefore, the profiles of the reference sample presented in Fig.~\ref{fig:XrayProfiles} and used in the calculations are the ones obtained without inclination of the sample ($\chi=0$), in contrast to our previous analysis \cite{kaganer12NWstrain} where profiles integrated over $\chi$ had to be employed.

The calculated x-ray diffraction profiles are also shown in Fig.~\ref{fig:XrayProfiles}. Integrations over the distributions of nanowire orientations, lengths, diameters, and distances between bundled nanowires were performed as a Monte Carlo average of the intensity as obtained by Eq.~(\ref{eq:25}). The simulations depicted in Fig.~\ref{fig:XrayProfiles} are obtained by convoluting the calculated profiles with the resolution functions of the respective reflections.

\begin{figure*}
\includegraphics[width=\textwidth]{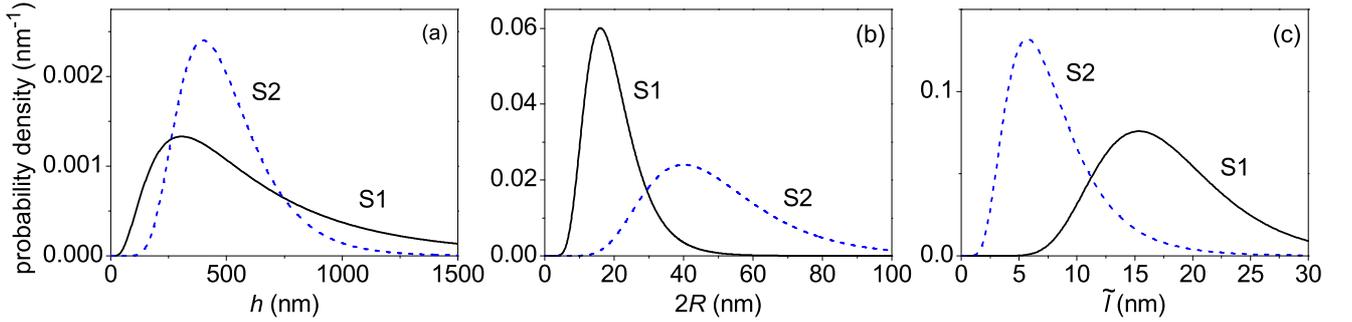} 
\protect\caption{Probability density distributions $\rho$ of (a) the nanowire lengths $h$, (b) the diameters $2R$, and (c) the effective distances $\tilde{l}$ used in the Monte Carlo calculation of the diffraction profiles.}
\label{fig:DistrParams}
\end{figure*}

The averages over the orientational distributions of the nanowires and these of the x-ray beams are modeled as follows. The scattering vector $\mathbf{K}$ in Eq.~(\ref{eq:A3}) is modeled by a Gaussian distribution of the angle $\psi$ describing the vertical divergence of the x-ray beam. Its full width at half maximum (FWHM) is taken to be 2.8$^{\circ}$, equal to that measured by tilting the bulk GaN sample \cite{kaganer12NWstrain}. Then, the scattering vector $\mathbf{q}$ is defined by Eq.~(\ref{eq:30}), where the nanowire orientation $\mathbf{n}$ is described by a Gaussian distribution of the nanowire tilt angle with an FWHM of 3.5$^{\circ}$, and a uniform distribution of the tilt azimuth from 0 to $2\pi$. The direction of the $x$ axis connecting two bundled nanowires is also taken to be uniformly distributed from 0 to $2\pi$.

With the scattering vector $\mathbf{q}$ thus defined, the calculation of the scattering amplitude given in Eq.~(\ref{eq:22}) requires the nanowire length $h$, its diameter $2R$, and the effective distance $\tilde{l}$ between two bundled nanowires. The lengths $h$ here are not the full lengths of the nanowires, which exceed 1~\textmu m, but the nanowire segments that bundle together, whose lengths are some hundreds of nanometers. Likewise, the diameters $2R$ are not the diameters of the bundled nanowires seen in the top-view scanning electron micrographs in Figs.~\ref{fig:SEMs}(b) and \ref{fig:SEMs}(d), but the smaller diameters of single nanowires forming these bundles. The length, diameter, and effective distance distributions in the real nanowire ensemble are correlated. First, thin as-nucleated nanowires experience bundling.  During further growth, these bundles with a larger total diameter and larger separation compared with the initial nanowires experience further bundling. However, in the present analysis we neglect these correlations for the sake of simplicity and assume the length, diameter, and effective distance distributions to be independent.

Figure \ref{fig:DistrParams} presents the distributions of the nanowire lengths $h$, diameters $2R$, and the effective distances $\tilde{l}$ used in the calculation of the diffraction profiles in Fig.~\ref{fig:XrayProfiles}. The choice of the distribution type for the nanowire lengths was found to be essential. Symmetric distributions, such as a Gaussian, do not lead to an agreement with the experiment. Asymmetric distributions, such as the log-normal distribution used here, enhance the weight of shorter nanowire segments as shown in Fig.~\ref{fig:DistrParams}(a) and result in a good agreement between experiment and theory. For simplicity, we use log-normal distributions for the diameters $2R$ and the effective distances $\tilde{l}$ as well, but the choice of the distribution type for these quantities is not crucial.

Figures \ref{fig:DistrParams}(a) and \ref{fig:DistrParams}(b) show that fairly broad distributions of the lengths of the nanowire segments and their diameters are needed to obtain an agreement of the calculated x-ray diffraction profiles and the experimental ones. For sample 1, the length distribution is characterized by $h=700\pm600$~nm, where the first number is the mean value and the second is the standard deviation.  For the diameter, the corresponding values are $2R=20\pm8$~nm for sample 1. For sample 2, we have used $h=500\pm200$~nm and $2R=50\pm20$~nm, respectively. The smaller lengths of the bundling segments and larger diameters for sample 2 agree with its larger coalescence degree evident from Fig.\ \ref{fig:SEMs}.

\begin{figure}[b]
\includegraphics[width=1\columnwidth]{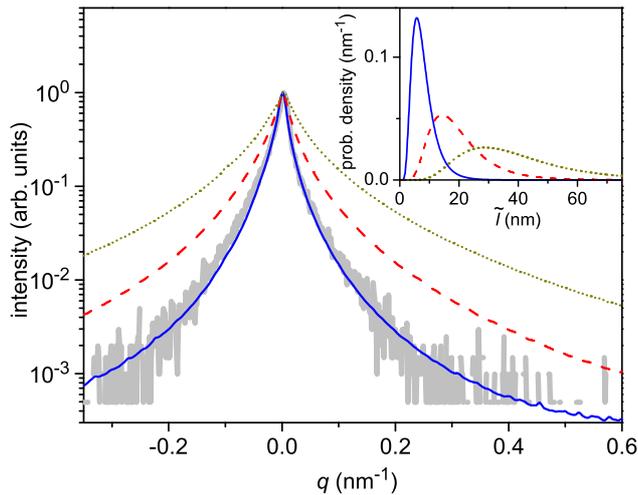} \protect\caption{X-ray diffraction profile of sample 2 across the 0002 reflection (thick gray line) and the profiles calculated for different distributions of the effective distance $\tilde{l}$. The length and diameter distributions are taken the same as used for sample 2 in Figs.~\ref{fig:DistrParams}(a) and \ref{fig:DistrParams}(b), respectively.}
\label{fig:NWdist}
\end{figure}

Deviations from these values degrade the quality of the fit of the experimental data by the simulated profiles. However, the most critical parameter for the shape of the calculated diffraction profiles is the effective distance $\tilde{l}$. The distances $\tilde{l}=18\pm6$~nm for sample 1 and $\tilde{l}=8\pm4$~nm for sample 2 are notably smaller than the average distance of about $l=30$~nm between the nucleated nanowires prior to their bundling. Already bundled nanowires that experience further bundling during their growth are separated by even larger distances $l$.  This result implies that a large fraction of the bending energy due to nanowire bundling is released by plastic deformation at the joints. On the other hand, these values of $\tilde l$ are notably larger than an estimate of $\tilde l$ from the energy minimization in Sec.\ \ref{sub:PlasticRelax}. Since introduction of dislocations between two merging atomically flat side facets of nanowires is hindered, the system may not reach its minimum of energy.

Let us estimate the dislocation densities in samples 1 and 2, using the mean values of the lengths, diameters, and effective distances obtained above. The estimate is crude since the distributions of these quantities are broad and their maxima occur at smaller values than their means. Besides, the distance between nanowires $l$ does not enter in the calculation of the x-ray diffraction profiles and has to be assigned separately. We take the mean distance between uncoalesced nanowires $l=30$~nm for the estimate. Then, the average number of dislocations at a joint as calculated by Eq.\ (\ref{eq:19a}) is $N\approx 1$ for sample 1 and $N\approx 7$ for sample 2. These numbers of dislocations per areas $2h \times 2R $ give rise to dislocation densities $4 \times 10^9$ and $1.4 \times 10^{10}$~cm$^{-2}$ for samples 1 and 2, respectively. Hence, larger coalescence degree of sample 2 gives rise to a larger dislocation density. In any case, this estimate shows that the dislocation densities are comparable with the nanowire densities.

Figure \ref{fig:NWdist} demonstrates the variation of the diffraction profiles when the mean effective distance $\tilde{l}$ is increased.  Calculations are performed for the same distributions of the lengths of the nanowire segments and the nanowire diameters as used for sample 2 in Figs.~\ref{fig:DistrParams}(a) and \ref{fig:DistrParams}(b), respectively. The solid blue line reproduces the calculated profile in Fig.~\ref{fig:XrayProfiles} and agrees with the experimental profile. The dotted line, in turn, is calculated for a distribution of $\tilde{l}$ possessing the maximum at $\tilde{l}=30$~nm. This distance between nanowires corresponds to a nanowire density of $4\times10^{10}$~cm$^{-2}$, which is a maximum density at the end of the nanowire nucleation process and before their massive coalescence \cite{kaganer16bundling}. Hence, this distribution of distances approximately corresponds to purely elastic bending of nanowires without plastic relaxation. Evidently, the profile obtained under this assumption is in gross disagreement with the experimental one.

\subsection{Line breadth analysis}

\begin{figure}
\includegraphics[width=0.9\columnwidth]{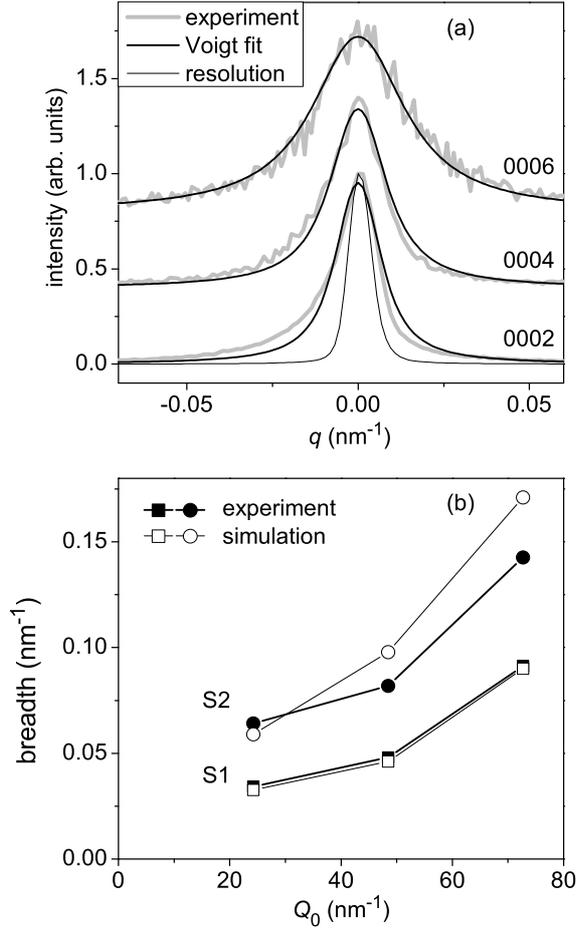}\protect\caption{(a) Voigt fits to the x-ray diffraction profiles of sample 1 and (b) integral breadths obtained from the experimental (full symbols) and the simulated (open symbols) profiles in different diffraction orders.}
\label{fig:VoigtFits}
\end{figure}

The comparison of the diffraction line widths in successive reflection orders as originally proposed by Williamson and Hall \cite{WilliamsonHall53} is an established tool of the size-strain analysis since decades\cite{warren:book69,birkholz06,guinebretiere:book} and has been used for GaN nanowire samples as well \cite{jenichen11,garrido14}.  The Williamson-Hall analysis is based on the assumption that the displacements of atoms are correlated only on short range, so that every small volume of the sample diffracts independently according to its local strain \cite{StokesWilson44}. Then, the broadening due to an inhomogeneous strain is proportional to the reflection order while the broadening due to finite sizes of the crystallites does not depend on the reflection order.

However, the x-ray scattering from a nanowire (or its segment) bent due to bundling as described by Eq.~(\ref{eq:22}) depends on the atomic displacements along the whole nanowire, and does not satisfy the assumptions underlying the Williamson-Hall analysis. As a result, the size and the strain effects cannot be resolved in a simple way.

Figure~\ref{fig:VoigtFits} presents the integral breadth analysis of samples 1 and 2. Since the widths of the diffraction lines themselves and the resolution functions are comparable, a deconvolution of the resolution is essential. We fit both the measured peaks and the resolution functions by Voigt functions [see Fig.~\ref{fig:VoigtFits}(a) as an example], and then subtract Gaussian and Lorentzian widths of the resolution functions. The integral breadths of the diffraction lines are calculated from the resolved Gaussian and Lorentzian widths (see, e.\,g., Ref.~\onlinecite{birkholz06}, Sec.\ 3.1). The same calculation is performed on the x-ray profiles calculated by the Monte Carlo method and presented in Fig.~\ref{fig:XrayProfiles}. 

The integral breadths for the measured and the calculated diffraction profiles are presented in Fig.~\ref{fig:VoigtFits}(b). The larger coalescence degree of sample 2 gives rise to broader diffraction peaks. However, non-linear dependencies of the integral breadths on the reflection order do not allow a clear distinction of size and strain effects. In fact, we cannot extract a single parameter from Eq.~(\ref{eq:22}) that would describe the slopes of the lines in Fig.~\ref{fig:VoigtFits}(b) and could serve as a mean-squared strain in the sense of the conventional Williamson-Hall analysis.

\section{Summary and conclusions\label{sec:Discussion}}

Our study suggests that the high-density ensembles of spontaneously formed GaN nanowires are not free of extended defects, but actually beset with dislocations. These defects originate from the overwhelming disposition of GaN nanowires to bundle after they have reached a certain critical length. This process is driven by the reduction of the surface energy of the side facets of the nanowires at the expense of the bending of their bottom segments. Our energy calculations predict that the elastic energy of bending can be reduced by the creation of dislocations at the joints. These dislocations possess comparatively small elastic energy since their strain fields are relaxed at distances exceeding the nanowire diameter due to the presence of free surfaces.

To elucidate the actual strain state of GaN nanowire ensembles, we have compared the experimental x-ray diffraction profiles from nanowire ensembles exhibiting substantial coalescence by bundling to calculated profiles assuming independent distributions of the lengths of the nanowire segments, their diameters, and the effective distances between nanowires that decribe their residual bending. The parameters yielding a good fit between the experiment and simulated x-ray diffraction profiles indicate that the bending is indeed partly accommodated by dislocations at the joints. The fact that the nanowires exhbit a residual bending suggests that plastic relaxation may be kinetically hindered by the difficulty to introduce dislocations when the atomically flat facets of the two nanowires meet. However, plastic relaxation at the joints may also induce the creation of large-angle boundaries between coalesced nanowire segments. The x-ray diffraction profiles studied in the present work are insensitive to the microstructure of the joints because of their small volume, and the available microcopic studies in the literature \cite{consonni09,jenichen11,grossklaus13,fan14} do not clarify this question either. Clearly, dedicated microscopic studies are needed to reveal the actual microstructure of the joints.

The dislocation density in typical GaN nanowire ensembles such as investigated in the present work is on the order of $10^{10}$~cm$^{-2}$, i.\,e., similar to the one encountered in heteroepitaxial GaN films grown by molecular beam epitaxy. However, contrary to the situation in films, the dislocations at the tilt boundary of bundled nanowires intersect the nanowire bundle in radial direction, and do not propagate along the nanowire axis. The effect of these dislocations is thus much less detrimental than that of threading dislocations intersecting a heteroepitaxial layer. Nevertheless, it is likely that these boundary dislocations act as nonradiative centers as well, and it would certainly be desirable to avoid their formation. Since the bundling of nanowires is inevitable once the nanowire lengths significantly exceed the distances between them, the only way to avoid this process is a drastic reduction of the nanowire nucleation density. This reduction, in turn, may be achieved either by selective area growth or by employing a substrate with structural and chemical properties that reduce the nucleation rate of GaN nanowires. 

\begin{acknowledgments} The authors thank Anne-Kathrin Bluhm for scanning electron microscopy and Javier Bartolomé Vilchez for a critical reading of the manuscript. 
\end{acknowledgments}


%

\end{document}